\documentclass{aa}
\usepackage{graphicx}
\usepackage{epsfig}
\usepackage{txfonts}
\usepackage{natbib}
\begin{document}

\title{Formation of current sheets and sigmoidal structure by the kink
       instability of a magnetic loop}

\author{B. Kliem    \inst{1}
   \and V. S. Titov \inst{2}
   \and T. T\"or\"ok\inst{1,3}}

\institute{Astrophysikalisches Institut Potsdam, 14482~Potsdam, Germany
      \and Theoretische Physik IV, Ruhr-Universit\"at Bochum, 44780 Bochum,
           Germany      
      \and School of Mathematics and Statistics, University of
           St\,Andrews, St\,Andrews, Fife KY16 9SS, UK}

\titlerunning{Sigmoids formed by kink unstable loop}

\offprints{B. Kliem, \email{bkliem@aip.de}}

\date{Received 4 April 2003 / Accepted 3 November 2003 } 

\abstract{We study dynamical consequences of the kink instability of a
twisted coronal flux rope, using the force-free coronal loop model by
\cite{Tit:Dem-99} as the initial condition in ideal-MHD simulations. When a
critical value of the twist is exceeded, the long-wavelength ($m\!=\!1$)
kink mode develops.
Analogous to the well-known cylindrical approximation, a helical
current sheet is then formed at the interface with the surrounding medium.
In contrast to the cylindrical case,
upward-kinking loops form a second, vertical current sheet below the loop
apex at the position of the hyperbolic flux tube (generalized X line) in
the model. 
The current density is steepened in both sheets and eventually exceeds the
current density in the loop
(although the kink perturbation starts to saturate in our simulations
without leading to a global eruption).
The projection of the field lines that pass
through the vertical current sheet shows an S shape whose sense agrees
with the typical sense of transient sigmoidal (forward or reverse S-shaped)
structures that brighten in soft X rays prior to coronal eruptions. The
upward-kinked loop has the opposite S shape, leading to the conclusion
that such sigmoids do not generally show the erupting loops themselves but
indicate the formation of the vertical current sheet below them that is the
central element of the standard flare model. 

}

\maketitle

\section{Introduction}
\label{intro}

Rising filaments, flares, and coronal mass ejections 
on the Sun often show the phenomenology of a loop-shaped magnetic flux
system with fixed footpoints at the coronal base and signatures of magnetic
twist. 
A single twisted magnetic flux rope appears to contain essential elements
of the magnetic topology of the unstable, often erupting flux. Since the
ratio of kinetic to magnetic pressure, the plasma beta, is very small in
the inner corona, $\beta\sim10^{-3}\dots10^{-2}$, the flux system must be
nearly force free in the quasi static pre-eruption state, i.e.,
\begin{equation}
\nabla\mbox{\boldmath$\times$}\vec{B}=\alpha(\vec{x})\vec{B}\,.
\label{eq_forcefree}
\end{equation}

\begin{figure}    
\begin{center}
 \resizebox{.85\hsize}{!}                
           {\includegraphics{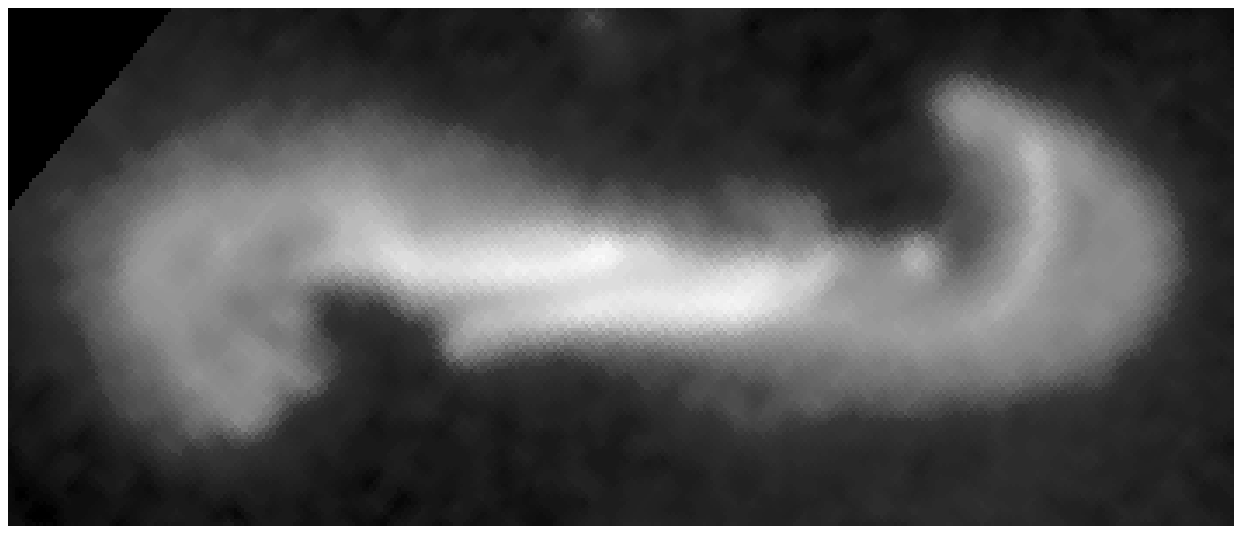}}
\end{center}
\vspace{-3mm}
\caption[]
{Transient sigmoid observed in an eruptive flare on 1994 Oct 25 
 (from \citeauthor{Mano:al-96}, \citeyear{Mano:al-96}) in AR 7792, 
 which had primarily left-handed twist \citep{Aura:al-99,vanDrie:al-00}.} 
\label{fig_TD99f1}
\end{figure}

Such major energy release events, which we will refer to collectively as
flares in the following, often show a distinctive soft X-ray (SXR)
brightening of sigmoidal (S or reverse-S) shape shortly prior or
during their early, impulsive stages of development
\cite[e.g.,][]
{Rus:Kum-96,Mano:al-96,Pevt:al-96,Aura:al-99,Moor:al-01,Gibs:al-02}. 
The sigmoidal shape is generally regarded as a signature of coronal
currents in a twisted field and shows a hemispheric preference: forward
(reverse) S shapes
dominate in the southern (northern) hemisphere \citep{Rus:Kum-96}.
Typically the central
part of such transient sigmoids is approximately aligned with the neutral
line of the normal photospheric field component, often also with a filament.
They are bright and more sharply defined in the middle, while their ends
fan out more diffusively at opposite sides of the neutral line. Sometimes
their middle section is slightly split, giving the structure the appearance
of two point-symmetric letters J. Each J may consist of several fibres
(Fig.\,\ref{fig_TD99f1}). 
%

The solar soft X-ray corona often also contains long-lived sigmoidal
structures that appear to form a separate class. These typically exist for
more than a day, do not show the concentration in the middle, may be
composed of many fibres that show the S shape collectively rather than
individually, and may form not only within but also between active regions.

The sense of the sigmoidal shape and the handedness of the magnetic twist
are correlated: \cite{Pevt:al-97} found that more than 90{\%} of all
sigmoids in those active regions that have a uniform sign of $\alpha$ in
the photospheric magnetogram show forward (reverse) S shape for
$\alpha>(<)~0$, i.e., for right (left) handed twist. We will refer to this
result as $\sigma$-$\alpha$ correlation, where $\sigma>(<)~0$ stands for
forward (reverse) S shape. The correlation is consistent
with the hemispheric preference, since $\alpha$ is predominantly $>(<)~0$ in
the southern (northern) hemisphere \citep{Seehafe-90,Pevt:al-95}.

The transient sigmoids that brighten in flares 
(both confined and eruptive) have so far mostly been interpreted as an
illumination of the unstable flux system itself. 
\cite{Rus:Kum-96} proposed that they show a sufficiently twisted,
kink-unstable flux rope. \cite{Pevt:al-96}, \cite{Moor:al-01}, and others
suggested further that the unstable flux rope acquired its supercritical
twist through the reconnection of two nearly aligned, subcritical flux
ropes near their end points, which conforms to the strong brightening in
the middle of the sigmoids and may symbolically be written as $J+J\to{S}$.
Transient sigmoids were thus regarded by \cite{Moor:al-01} as supporting
their ``tether cutting'' model of flares 
(although they noted that another, much fainter structure was moving in
sequences of SXR images in some of their events). 

The presence of a SXR sigmoid correlates with the likelihood of a flaring
active region
to cause an eruption \citep{Canf:al-99}. A ``standard'' model for the
main phase of eruptive flares is now widely accepted
\cite[e.g.,][]{Shibata-99}. It includes the formation of a large-scale
vertical current sheet below the erupting flux, in which magnetic
reconnection leads to the long-lasting energy release. This
model is assumed to hold independently of the question whether the event is
initiated by an instability of a single flux rope, by reconnection between
flux ropes (tether cutting), or by other mechanisms, e.g., ``magnetic
breakout'' \citep{Anti:al-99}. 

The canonical instability of a twisted flux rope is the kink instability.
In applications to the solar corona,
the simplifying assumption of straight, cylindrically symmetric flux ropes
has always been used so far \cite[e.g.,][]{Hood-92}. This is
based on the typically large aspect ratio
of solar magnetic loops. The instability occurs if the twist, 
$\Phi\!=\!lB_\phi(r)/(rB_z(r))$, exceeds a critical value,
$\Phi_\mathrm{c}$. Here $l$ is the length of the flux rope, $r$ is
the (minor) radius, and $B_z$ and $B_\phi$ are the axial and azimuthal
field components, respectively. The critical value depends on the details
of the considered equilibrium. 

In this paper we address the relationship between the kink instability of
an arched twisted flux rope and the formation of current sheets and
sigmoidal structures, starting from the analytical magnetic loop
equilibrium by \cite{Tit:Dem-99}, cited as T\&D in the following (see their
Fig.\,2 for a schematic). This approximate, cylindrically symmetric,
force-free equilibrium consists of a toroidal ring current of
major radius $R$ and
minor radius $a$, whose outward-directed Lorentz self-force is balanced
with the help of a field by two fictitious magnetic charges of opposite sign
which are placed at the symmetry axis of the torus at distances ${\pm}L$ to
the torus plane. That axis lies below the photospheric plane $\{z\!=\!0\}$
at a depth $d$. The resulting field outside the torus is current-free and
contains a concentric magnetic X line between the torus and its centre. A
toroidal field component created by a fictitious line current running along
the axis of symmetry is added. This results in a force-free field inside
the torus with the twist taking a finite value everywhere in the
configuration, whose part in the volume $\{z\!>\!0\}$ is a model of a
coronal loop (or flux rope). 

The presence of a toroidal field component turns the neighbourhood of
the X line into a hyperbolic flux tube (HFT) which consists of two 
intersecting quasi-separatrix layers with extremely diverging field lines
(for a strict definition of HFTs, see \citeauthor{Tito:al-02},
\citeyear{Tito:al-02}). The existence of the HFT is generic to
such force-free loop configurations with a nonvanishing net current 
(\citeauthor{Tito:al-99}, \citeyear{Tito:al-99}), which is important for
understanding the origin of sigmoidal structures in twisted
configurations.

We present two characteristic cases of the kink instability, which show the
different possibilities of current sheet and sigmoid formation in the T\&D
equilibrium, leaving a systematic study of the instability for a separate
paper \citep{Toer:al-03}. 

T\&D pointed out that the torus is also unstable with respect to global
expansion (growing perturbations ${\delta}R>0$). This has recently been
confirmed by \cite{Rous:al-03}. 
We expect that the global expansion instability leads to current sheets and
sigmoidal structures in a manner largely analogous to the upward kink
instability.


We integrate the compressible ideal MHD equations using the simplifying
assumption $\beta=0$ which has no influence on the qualitative evolution of
the kink instability in its linear phase, during which the current sheets
and sigmoids are formed: 
\begin{eqnarray}
\partial_t\rho&=&
                 -\nabla\cdot(\rho\,\vec{u})\,,           \label{eq_rho}\\
\rho\,\partial_{t}\vec{u}&=&
      -\rho\,(\,\vec{u}\cdot\nabla\,)\,\vec{u}
      +\vec{j}\mbox{\boldmath$\times$}\vec{B} \,,         \label{eq_mot}\\
\partial_{t}\vec{B}&=& 
    \nabla\mbox{\boldmath$\times$}(\,\vec{u}\mbox{\boldmath$\times$}
    \vec{B}\,)\,,                                         \label{eq_ind}\\
\vec{j}&=&\mu_0^{-1}\nabla\mbox{\boldmath$\times$}\vec{B}\,. \label{eq_cur}
\end{eqnarray}

The equilibrium by T\&D is used as the initial condition for the field,
with the loop chosen to lie in the plane $\{x\!=\!0\}$. The initial density
distribution is specified such that the Alfv\'en velocity is uniform,
$\rho_0=B_0^2/\mu_0$. Lengths, velocities, and times are normalized,
respectively, by $L$, the initial Alfv\'en velocity,
$v_\mathrm{a0}=B_0/(\mu_0\rho_0)^{1/2}$, and the 
Alfv\'en time, $\tau_\mathrm{a}=L/v_\mathrm{a0}$. 

A modified Lax-Wendroff scheme is employed on a nonuniform Cartesian grid
in a ``half cube''
$[-L_x,L_x]\times[0,L_y]\times[0,L_z]$
with closed boundaries except the front boundary, $\{y\!=\!0\}$, where
line symmetry with respect to the $z$ axis is implemented. The latter
implies $u_{x,y}(0,0,z)=0$ so that only vertical displacements of
the loop apex are permitted. Viscosity is included in Eq.\,(\ref{eq_mot}) to
improve numerical stability. The numerical details are the same as in
\cite{Toer:al-03}.

\section{Kink instability}
\label{kink}

\begin{figure}    
\resizebox{\hsize}{!}                
          {\includegraphics{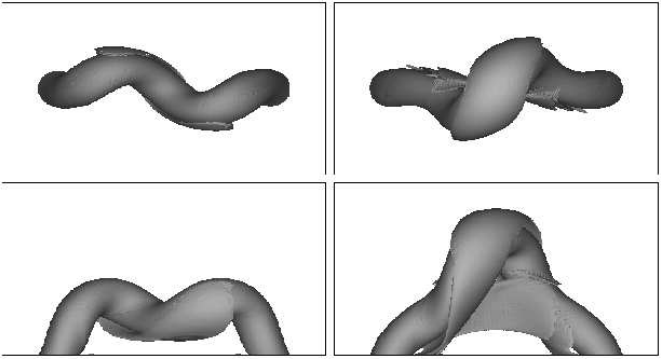}}
\caption[]
{Top and side view of current density isosurfaces for 
 $\Phi=4.9\pi$.
 \emph{Left}:  $|\vec{j}|=0.15\,j_{\rm max}$ at $t=35$; unperturbed case. 
 \emph{Right}: $|\vec{j}|=0.25\,j_{\rm max}$ at $t=28$; run with an
                upward initial velocity perturbation at the loop apex. 
 $|x|\le1.5$, $|y|\le3$, $0\le z\le3$.} 
\label{fig_iso_j_Nt=15}
\end{figure}

The parameters of the basic loop equilibrium are chosen similarly to T\&D:
$d=L=1$ (50~Mm) and $R=2.2$. The twist is specified via the number of
turns that the field lines at the surface of the torus make about its axis
(counted for the whole torus), taken here to be $N_\mathrm{t}=15$. This
also fixes the minor radius of the loop, $a=0.32$. The
resulting twist of the coronal part of
the loop, averaged over its cross section, $\Phi=4.9\pi$, is about
$1.4\Phi_\mathrm{c}$ of this configuration \citep{Toer:al-03}. Here the
long-wavelength ($m\!=\!1$) kink mode develops spontaneously with downward
displacement of the loop apex; it is initiated by the weak,
downward-directed forces that result from the initial discretization errors
of the current density on the grid. The displacement and velocity of the
apex rise exponentially until the apex height is reduced to $\approx2/3$ of
its initial value
by $t\approx25$, after which the instability starts to saturate. The
configuration that results for left-handed twist ($\alpha<0$) is shown in
Fig.\,\ref{fig_iso_j_Nt=15} (left) by current density isosurfaces. The kink
perturbation forms a helical current sheet that is wrapped around the
displaced parts of the loop at the interface with the surrounding plasma in
a manner similar to the cylindrical case. The current density in this sheet
rises exponentially as well, eventually exceeding the current density in
the loop. The projection of both the loop and the current sheet onto the
bottom plane possess a reverse S shape, as is characteristic of sigmoids in
left-handed twisted fields. If the handedness of the twist is reversed,
then the $m\!=\!1$ kink mode develops with identical time profile but
reversed handedness of the loop perturbation, i.e., forward S shape of its
photospheric projection. 
%

In order to study the kink mode with upward apex displacement, a small
upward initial velocity perturbation is applied at the apex. It is ramped
up over $5\tau_\mathrm{a}$ to $u_z=0.01$, then switched off. In this run
the $m\!=\!1$ kink mode develops with a similar growth rate as before,
but again starts to saturate after
$t\approx25$, at an apex velocity of $u_z\approx0.1$. 
The resulting configuration in the case of left-handed twist is also shown
in Fig.\,\ref{fig_iso_j_Nt=15} (right). In addition to the helical current
sheet that is again formed as the kink perturbation piles up the
surrounding flux, a second current sheet, which has no analogue in the
cylindrical case (but corresponds to the main element of the standard flare
model), now occurs below the rising section of the loop. 
It is formed by a pinching of the HFT in the lateral stagnation flow set up
below the rising loop apex. For a detailed analysis of the pinching
mechanism, which is a three-dimensional generalization of the well-known
pinching of X points in the 2D case, we refer to
\cite{TGN:al-03} and \cite{GTN:al-03}. 
The current density rises exponentially in both sheets, exceeding
that in the kinked loop in the nonlinear stage of the instability. The
rising unstable loop and the helical current sheet now both possess a
forward S-shaped photospheric projection -- opposite to the characteristic
sigmoid shape in left-handed fields.
Reversing the twist of the initial configuration
leads to the upward $m\!=\!1$ kink mode with identical time profile but
reversed handedness of the kink perturbation. It should be noted that the
applied velocity perturbation is exactly vertical and so does not introduce
a preference for the handedness of the kink perturbation.
The two configurations of opposite sigmoidal shape shown in
Fig.\,\ref{fig_iso_j_Nt=15} possess the same photospheric $\alpha$.

\section{Formation of sigmoids}
\label{sigmoids}

In all four cases discussed in the previous section, the handedness of the
kink perturbation of the loop axis (sign of ``external writhe'') equals the
handedness of the field line twist (sign of ``internal writhe''). This is a
general property of the kink instability \cite[known as resonance
at the point of marginal stability; see, e.g.][]{Lint:al-99}. 
Its immediate consequence is that \emph{transient sigmoids cannot be
identified with rising kink-unstable loops}. 
Since there are strong observational indications for the existence of
substantial magnetic twist in eruptive flux systems, 
transient sigmoids must in general be associated with another structure in
the unstable twisted flux. 

\begin{figure}    
\resizebox{\hsize}{!}                
          {\includegraphics{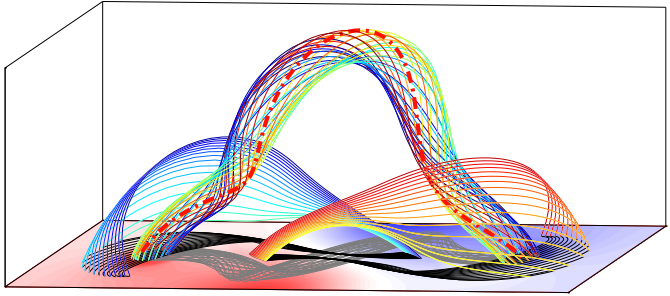}}
\caption[]
{Field lines of the configuration shown in the right panels of
 Fig.\,\ref{fig_iso_j_Nt=15}. $|x|\le0.8$, $|y|\le3$, $0\le z\le2.2$. Loop
 field lines lie in the surface $r=a/2$ (to separate the loop clearly
 from the sigmoids); one of them is emphasized to
 indicate the left handedness of the twist. The remaining field lines pass
 through two symmetrical vertical stripes close to the $z$ axis that bracket
 the pinched HFT. The normal
 component of the magnetic field in the bottom plane is shown color-coded
 (blue -- positive, red -- negative). Projections of the field lines are
 overlayed.} 
\label{fig_jjsigma}
\end{figure}

\begin{figure}    
\resizebox{\hsize}{!}                
          {\includegraphics{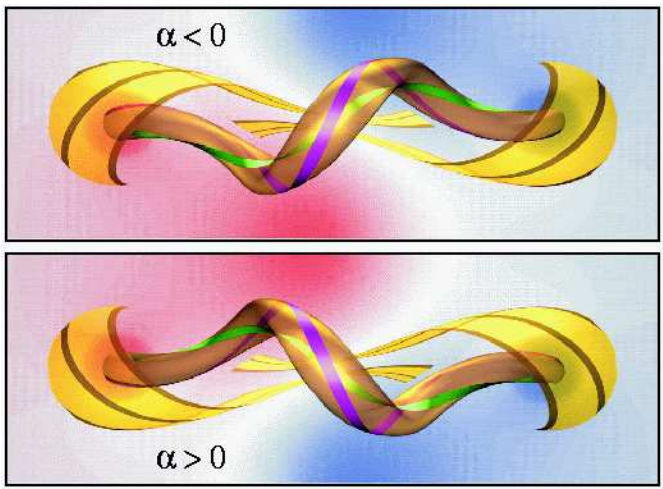}}
\caption[]
{Top view on selected field lines and on the surfaces that they form.
 $|x|\le1$, $|y|\le3$. 
 The field lines for $\alpha<0$ are identical to those plotted in
 Fig.\,\ref{fig_jjsigma}. The bottom panel shows similarly selected field
 lines of the system with reversed $\alpha$ at the same time.} 
\label{fig_KJplt}
\end{figure}

We follow the suggestion by T\&D that bright SXR structures outline magnetic
structures with enhanced current density, and hence enhanced dissipation,
and calculate field lines that pass through the vertical current sheet
below the upward-rising kinked loop 
(Figs.\,\ref{fig_jjsigma} and
\ref{fig_KJplt}). These field lines form a sigmoidal surface that not only
follows the $\sigma$-$\alpha$ correlation
but also corresponds nicely to further properties of transient sigmoids
like the near-alignment with the photospheric neutral line, the
concentration in the middle and the diffuse spreading at the ends. The
occasionally observed splitting 
into two J-shaped parts, which may themselves consist of multiple J-shaped
fibres, can readily be explained by assuming the existence
of multiple current density peaks within the current sheet. This can
naturally be expected in a fibrous corona (we have indicated this effect in
Figs.\,\ref{fig_jjsigma} and \ref{fig_KJplt} by selecting field lines that
pass closely to the $z$ axis; in our simulation of a smooth initial
configuration, the current density peaks
exactly at the $z$ axis). Flux pileup in the stagnation flow at both sides
of the current sheet may also lead to the effect. Finally, cases of sigmoid
lengthening, as described by \cite{Mano:al-96}, are consistent with the new
interpretation only. 

Magnetic reconnection is expected to occur in the vertical current sheet,
amplifying the brightening of the sigmoidal field lines at SXR and
permitting transitions between continuous (S) and broken (double J)
sigmoidal patterns (in both directions). 

With this interpretation, transient sigmoids cannot be considered as direct
support of the tether cutting model \citep{Moor:al-01} but, of course, they
do not contradict the model in any manner. Our
interpretation is not
undermined by the fact that the kink-unstable loop does
not succeed to evolve into a true eruption in our simulations
\citep{Toer:al-03}, because transient sigmoids occur at the onset of
eruptive events. 


The downward-kinked loop may be associated with long-lived sigmoids. As the
instability saturates, this kink perturbation may lead to a stable
structure which follows the $\sigma$-$\alpha$ correlation and is
nearly aligned in the middle with the photospheric neutral line. The dip in
the loop or in the flux below it supports the formation of a filament,
which is often observed in close spatial association with sigmoids
\citep{Pevtsov-02}. Further studies are required to check this conjecture. 
See \cite{Toe:Kli-03} for an alternative interpretation of long-lived
sigmoids as loops of enhanced current density but subcritical twist.

\section{Conclusions}
\label{concl}

The magnetic loop equilibrium by T\&D is unstable against the
long-wavelength ($m\!=\!1$) ideal kink mode if the magnetic twist exceeds a
critical value. The handedness of the
kink perturbation equals the handedness of the field line twist.
Consequently, the photospheric projection of rising kink-unstable loops
develops a forward (reverse) S shape for left (right) handed twist. This
is opposite to the strong observed ($\sigma$-$\alpha$) correlation of
sigmoid shape with magnetic handedness. Hence,
transient SXR sigmoids, which brighten in major energy release events,
preferably in eruptive events, cannot in general show the presumed
kink-unstable flux itself. We suggest that transient
sigmoids outline field lines which pass through the current density
enhancement in a vertical current sheet formed below the rising loop. This
interpretation 
leads to agreement with the $\sigma$-$\alpha$ correlation (and
hemispheric preference) as well
as with their alignment with the photospheric neutral line and with their
further specific geometric properties. Transient sigmoids thus indicate the
formation of the vertical current sheet that is the central element of the
standard model for eruptive flares. The question under which conditions a
rising kink-unstable loop develops into an eruption requires further
study. 

Downward-kinking loops take a sigmoidal shape that follows the observed
correlation with handedness. They may be related to long-lived sigmoids
and to filament formation.

\begin{acknowledgements}
We thank R. J. Leamon for pointing out the relationship of our calculations
to the model by \cite{Rus:Kum-96}. 
This investigation was supported by BMBF/DLR grants No.\ 50\,OC\,9901\,2
and 01\,OC\,9706\,4, by the Volkswagen Foundation, and by EU grant No.\
HPRN-CT-2000-00153. The John von Neumann-Institut f\"ur Computing, J\"ulich
granted Cray computer time.
\end{acknowledgements}


\begin{thebibliography}{}
\bibitem[Antiochos et al.(1999)Antiochos, DeVore, \& Klimchuk]{Anti:al-99}
         Antiochos, S. K., DeVore, C. R., \& Klimchuk, J. A. 1999, ApJ,
         510, 485
\bibitem[Aurass et al.(1999)]{Aura:al-99}
         Aurass, H., Vr\v{s}nak, B., Hofmann, A., \& Rud\v{z}jak, V. 1999,
	 Sol.\ Phys., 190, 267 
\bibitem[Canfield et al.(1999)Canfield, Hudson, \& McKenzie]{Canf:al-99}
         Canfield, R. C., Hudson, H. S., \& McKenzie, D. E. 1999, Geophys.
	 Res. Lett., 26, 627
\bibitem[Galsgaard et al.(2003)Galsgaard, Titov, \& Neukirch]{GTN:al-03}
         Galsgaard, K., Titov, V. S., \& Neukirch, T. 2003, ApJ, 595, 506
\bibitem[Gibson et al.(2002)]{Gibs:al-02} 
         Gibson, S.~E.~et al.\ 2002, ApJ, 574, 1021
\bibitem[Hood(1992)]{Hood-92}
         Hood, A. W. 1992, Plasma Phys.\ \& Contr.\ Fusion, 34, 411
\bibitem[Linton et al.(1999)]{Lint:al-99}
         Linton, M. G., Fisher, G. H., Dahlburg, R. B., \& Fan, Y. 1999,
	 ApJ, 522, 1190
\bibitem[Manoharan et al.(1996)]{Mano:al-96}
         Manoharan, P. K., van Driel-Gesztelyi, L., Pick, M., \& D\'emoulin,
	 P. 1996, ApJ, 468, L73
\bibitem[Moore et al.(2001)]{Moor:al-01}
         Moore, R.~L., Sterling, A.~C., Hudson, H.~S. \& Lemen, J.~R. 2001,
	 ApJ, 552, 833 
\bibitem[Pevtsov(2002)]{Pevtsov-02}
         Pevtsov, A. A. 2002, Sol.\ Phys., 207, 111
\bibitem[Pevtsov et al.(1995)Pevtsov, Canfield, \& Metcalf]{Pevt:al-95}
         Pevtsov, A.A., Canfield, R.C., \& Metcalf, T.R. 1995, ApJ,
         440, L109
\bibitem[Pevtsov et al.(1996)Pevtsov, Canfield, \& Zirin]{Pevt:al-96}
         Pevtsov, A.A., Canfield, R.C., \& Zirin. H. 1996, ApJ, 473, 533
\bibitem[Pevtsov et al.(1997)Pevtsov, Canfield, \& McClymont]{Pevt:al-97}
         Pevtsov, A. A., Canfield, R. C., \& McClymont, A. N. 1997, ApJ,
	 481, 973
\bibitem[Roussev et al.(2003)]{Rous:al-03}
         Roussev, I. I., et al. 2003, ApJ, 588 L45
\bibitem[Rust \& Kumar(1996)]{Rus:Kum-96}
         Rust, D. M., \& Kumar, A. 1996, ApJ, 464, L199
\bibitem[Seehafer(1990)]{Seehafe-90}
         Seehafer, N. 1990, Sol.\ Phys., 125, 219
\bibitem[Shibata(1999)]{Shibata-99}
         Shibata, K. 1999, Ap{\&}SS, 264, 129
\bibitem[Titov \& D\'emoulin(1999)]{Tit:Dem-99}
         Titov, V. S., \& D\'emoulin, P. 1999, A{\&}A, 351, 707 (T\&D)
\bibitem[Titov et al.(1999)Titov, D\'emoulin \& Hornig]{Tito:al-99}
         Titov, V. S., D\'emoulin, P., \& Hornig, G. 1999, in Magnetic Fields
         \& Solar Processes, ed. A. Wilson (ESA SP-448) 715
\bibitem[Titov et al.(2003)Titov, Galsgaard, \& Neukirch]{TGN:al-03}
         Titov, V.S., Galsgaard, K., \& Neukirch, T. 2003, ApJ, 582, 1172 
\bibitem[Titov et al.(2002)Titov, Hornig, \& D\'emoulin]{Tito:al-02}
         Titov, V. S., Hornig, G., \& D\'emoulin, P. 2002, J. Geophys. Res.,
	 107, doi:10.1029/2001JA000278 
\bibitem[T\"or\"ok \& Kliem(2003)]{Toe:Kli-03}
         T\"or\"ok, T., \& Kliem, B. 2003, A{\&}A, 406, 1043
\bibitem[T\"or\"ok et al.(2003)T\"or\"ok, Kliem \& Titov]{Toer:al-03}
         T\"or\"ok, T., Kliem, B., \& Titov, V. S. 2003, A{\&}A, in press
\bibitem[van Driel-Gesztelyi et al.(2000)]{vanDrie:al-00}
         van Driel-Gesztelyi, L., Manoharan, P. K., D\'emoulin, P., et al.
	 2000, J. Atmospheric \& Solar Terr. Phys., 62, 1437
\end{thebibliography}
\end{document}